\input harvmac\skip0=\baselineskip

\newcount\figno
\figno=0
\def\fig#1#2#3{
\par\begingroup\parindent=0pt\leftskip=1cm\rightskip=1cm\parindent=0pt
\baselineskip=11pt \global\advance\figno by 1 \midinsert
\epsfxsize=#3 \centerline{\epsfbox{#2}} \vskip 12pt {\bf Fig.\
\the\figno: } #1\par
\endinsert\endgroup\par
}
\def\figlabel#1{\xdef#1{\the\figno}}
\def\encadremath#1{\vbox{\hrule\hbox{\vrule\kern8pt\vbox{\kern8pt
\hbox{$\displaystyle #1$}\kern8pt} \kern8pt\vrule}\hrule}}

\def\p{\partial}



\lref\aes{ A.~Strominger, ``Macroscopic Entropy of $N=2$ Extremal
Black Holes,'' Phys.\ Lett.\ B {\bf 383}, 39 (1996)
[arXiv:hep-th/9602111].
}

\lref\ascv{ A.~Strominger and C.~Vafa, ``Microscopic Origin of the
Bekenstein-Hawking Entropy", Phys.\ Lett.\ B {\bf 379}, 99 (1996)
[arXiv:hep-th/9601029]. }

\lref\bmpv{ J.C. Breckenridge, R.C. Myers, A.W. Peet and C. Vafa,
``D-branes and Spinning Black Holes", hep-th/9602065. }

\lref\bmss{ R.~Britto-Pacumio, A.~Maloney, M.~Stern and
A.~Strominger, ``Spinning bound states of two and three black
holes,'' JHEP {\bf 0111}, 054 (2001) [arXiv:hep-th/0106099].
}

\lref\bpsb{ M. Marino, R. Minasian, G. Moore and A. Strominger,
``Nonlinear Instantons from Supersymmetric $p$-Branes",
hep-th/9911206. }

\lref\bsv{ R.~Britto-Pacumio, A.~Strominger and A.~Volovich,
``Two-black-hole bound states,'' JHEP {\bf 0103}, 050 (2001)
[arXiv:hep-th/0004017].
}

\lref\claus{ P.~Claus, M.~Derix, R.~Kallosh, J.~Kumar,
P.~K.~Townsend and A.~Van Proeyen, ``Black holes and
superconformal mechanics,'' Phys.\ Rev.\ Lett.\  {\bf 81}, 4553
(1998) [arXiv:hep-th/9804177].}

\lref\DabhBY{
  A.~Dabholkar, F.~Denef, G.~W.~Moore and B.~Pioline,
  ``Exact and asymptotic degeneracies of small black holes,''
  arXiv:hep-th/0502157.
}

\lref\dff{ V.~de Alfaro, S.~Fubini and G.~Furlan, `Conformal
Invariance In Quantum Mechanics,'' Nuovo Cim.\ A {\bf 34}, 569
(1976). }

\lref\dzero{ M. Bill\'o, S. Cacciatori, F. Denef, P. Fr\'e, A. Van
Proeyen and D. Zanon, ``The 0-brane action in a general D=4
supergravity background'', hep-th/9902100. }

\lref\fks{ S.~Ferrara, R.~Kallosh and A.~Strominger, ``N=2
extremal black holes,'' Phys.\ Rev.\ D {\bf 52}, 5412 (1995)
[arXiv:hep-th/9508072].
}

\lref\fvijay{V.~Balasubramanian and F.~Larsen, ``On D-Branes and
Black Holes in Four Dimensions,'' Nucl.\ Phys.\ B {\bf 478}, 199
(1996) [arXiv:hep-th/9604189].}

\lref\gghpr{J.~P.~Gauntlett, J.~B.~Gutowski, C.~M.~Hull, S.~Pakis
and H.~S.~Reall, ``All Supersymmetric Solutions of Minimal
Supergravity in Five Dimensions,'' arXiv:hep-th/0209114.}

\lref\gqkahler{ N. Reshetikhin and L. Takhtajan, ``Deformation
Quantization of K\"ahler Manifolds'', math.QA/9907171. }

\lref\GregoryTE{
  R.~Gregory, J.~A.~Harvey and G.~W.~Moore,
  Adv.\ Theor.\ Math.\ Phys.\  {\bf 1}, 283 (1997)
  [arXiv:hep-th/9708086].
}

\lref\gssy{D.~Gaiotto, A.~Simons, A.~Strominger and X.~Yin,
``D0-branes in Black Hole Attractors,'' arXiv:hep-th/0412179.}

\lref\gt{ G.~W.~Gibbons and P.~K.~Townsend, ``Black holes and
Calogero models,'' Phys.\ Lett.\ B {\bf 454}, 187 (1999)
[arXiv:hep-th/9812034].
}

\lref\gv{R.~Gopakumar and C.~Vafa, ``M-theory and Topological
Strings - I,II", hep-th/9809187; hep-th/9812127.}

\lref\iceland{ R.~Britto-Pacumio, J.~Michelson, A.~Strominger and
A.~Volovich, ``Lectures on superconformal quantum mechanics and
multi-black hole  moduli spaces,'' arXiv:hep-th/9911066.
}

\lref\jmjmas{\ J.~M.~Maldacena, J.~Michelson and A.~Strominger,
``Anti-de Sitter fragmentation,'' JHEP {\bf 9902}, 011 (1999)
[arXiv:hep-th/9812073]. }

\lref\juan{ J.~M.~Maldacena, `The large N limit of superconformal
field theories and supergravity,'' Adv.\ Theor.\ Math.\ Phys.\
{\bf 2}, 231 (1998) [Int.\ J.\ Theor.\ Phys.\  {\bf 38}, 1113
(1999)] [arXiv:hep-th/9711200]. }

\lref\juanbh{J.~M.~Maldacena, ``N = 2 extremal black holes and
intersecting branes,'' Phys.\ Lett.\ B {\bf 403}, 20 (1997)
[arXiv:hep-th/9611163].}

\lref\kall{R.~Kallosh, A.~Rajaraman, W.~K.~Wong, ``Supersymmetric
Rotating Black Holes and Attractors", Phys.\ Rev.\ D {\bf 55},
3246 (1997) [arXiv:hep-th/9611094].}

\lref\kkv{S.~Katz, A.~Klemm and C.~Vafa, ``M-Theory, Topological
Strings and Spinning Black Holes", Adv.\ Theor.\ Math.\ Phys.\
{\bf 3}, 1445 (1999) [arXiv:hep-th/9910181]. }

\lref\MohauptMJ{ T.~Mohaupt, ``Black hole entropy, special
geometry and strings,'' Fortsch.\ Phys.\  {\bf 49}, 3 (2001)
[arXiv:hep-th/0007195]. }

\lref\msas{M.~Spradlin and A.~Strominger, ``Vacuum states for
AdS(2) black holes,'' JHEP {\bf 9911}, 021 (1999)
[arXiv:hep-th/9904143].}

\lref\msone{ J.~Michelson and A.~Strominger, ``The geometry of
(super)conformal quantum mechanics,'' Commun.\ Math.\ Phys.\  {\bf
213}, 1 (2000) [arXiv:hep-th/9907191].
}

\lref\mss{ A.~Maloney, M.~Spradlin and A.~Strominger,
``Superconformal multi-black hole moduli spaces in four
dimensions,'' JHEP {\bf 0204}, 003 (2002) [arXiv:hep-th/9911001].
}

\lref\mstwo{ J.~Michelson and A.~Strominger, ``Superconformal
multi-black hole quantum mechanics,'' JHEP {\bf 9909}, 005 (1999)
[arXiv:hep-th/9908044].
}

\lref\msw{ J. Maldacena, A. Strominger and E. Witten, ``Black Hole
Entropy in M-Theory", hep-th/9711053. }

\lref\myers{ R.~C.~Myers, ``Dielectric-branes,'' JHEP {\bf 9912},
022 (1999) [arXiv:hep-th/9910053].
}

\lref\myersRV{ R.~C.~Myers, ``Nonabelian Phenomena on D-branes'',
Class.Quant.Grav.20, S347-S372 (2003) [arXiv:hep-th/0303072]. }

\lref\orv{ A.~Okounkov, N.~Reshetikhin and C.~Vafa, ``Quantum
Calabi-Yau and classical crystals,'' arXiv:hep-th/0309208.
}

\lref\osv{ H.~Ooguri, A.~Strominger and C.~Vafa, ``Black hole
attractors and the topological string,'' arXiv:hep-th/0405146.
}

\lref\rgcv{ R.~Gopakumar and C.~Vafa, ``M-theory and topological
strings. I,'' arXiv:hep-th/9809187. }

\lref\shmakova{ M.~Shmakova, `Calabi-Yau black holes,'' Phys.\
Rev.\ D {\bf 56}, 540 (1997) [arXiv:hep-th/9612076].
}

\lref\SimonsNM{ A.~Simons, A.~Strominger, D.~M.~Thompson and
X.~Yin, ``Supersymmetric branes in AdS(2) x S**2 x CY(3),''
arXiv:hep-th/0406121.
}

\lref\spinor{T. Mohaupt, ``Black Hole Entropy, Special Geometry
and Strings", hep-th/0007195.}

\lref\ssty{ A.~Simons, A.~Strominger, D.~M.~Thompson and X.~Yin,
``Supersymmetric branes in AdS(2) x S**2 x CY(3),''
arXiv:hep-th/0406121. }

\lref\suss{ B.~Freivogel, L.~Susskind and N.~Toumbas, ``A two
fluid description of the quantum Hall soliton,''
arXiv:hep-th/0108076.
}

\lref\VafaGR{ C.~Vafa, ``Black holes and Calabi-Yau threefolds,''
Adv.\ Theor.\ Math.\ Phys.\  {\bf 2}, 207 (1998)
[arXiv:hep-th/9711067].
}

\lref\VerCK{
  E.~Verlinde,
  ``Attractors and the holomorphic anomaly,''
  arXiv:hep-th/0412139.
}

\lref\gmt{
  J.~P.~Gauntlett, R.~C.~Myers and P.~K.~Townsend,
  ``Black holes of D = 5 supergravity,''
  Class.\ Quant.\ Grav.\  {\bf 16}, 1 (1999)
  [arXiv:hep-th/9810204].
}

\Title{\vbox{\baselineskip12pt\hbox{hep-th/0503217} }}{{ New
Connections Between 4D and 5D Black Holes}}

\centerline{Davide Gaiotto\footnote{*}{Permanent address:
Jefferson Physical Laboratory, Harvard University, Cambridge, MA,
USA.} ,~ Andrew Strominger* and Xi Yin* }
\smallskip\centerline{Center of Mathematical Sciences}
\centerline{ Zhejiang University, Hangzhou 310027 China}

\centerline{} \vskip.6in \centerline{\bf Abstract} { A simple
equality is proposed between the BPS partition function of a
general 4D IIA Calabi-Yau black hole and that of a 5D spinning
M-theory Calabi-Yau black hole.  Combining with recent results
then leads to a new relation between the 5D spinning BPS black
hole partition function and the square of the $N=2$ topological
string partition function.

 } \vskip.3in

\smallskip
\Date{}

\listtoc \writetoc

\newsec {Introduction}
Investigations of BPS black hole in string theory has shown them
to be a gold mine for deep and surprising physical and
mathematical insights.  In this paper we continue these
investigations in proposing and giving evidence for  a simple and
direct connection between a certain BPS partition function
$Z_{5D}$ of the general 5D spinning BPS black hole in a Calabi-Yau
compactification of M-theory and $Z_{4D}$  of the general 4D BPS
black hole in a Calabi-Yau compactification of the IIA theory.
Invoking prior results \osv\ then leads to a simple non-linear
relation between $Z_{5D}$ and topological string partition
function  $Z_{top}$.

We begin in section 2 by deriving the basic 4D-5D connection.
Exact 5D supersymmetric solutions were found in \gghpr\ which can
be described as a 5D black hole with $SU(2)_L$ spin
 $J^3_L$
and M2 charges $q^{5D}_A$ sitting at the center of a charge $p^0$
Taub-NUT. Since Taub-NUT is locally asymptotic to flat $R^3\times
S^1$ this implements a $5\to 4$ compactification. When the
compactification radius $R$, a modulus of the Taub-NUT solution,
becomes small the 4D picture becomes appropriate. We show that in
the 4D picture we have a black hole with D6-D2-D0 charges
$(p^0,{q^{5D}_A\over p^0}, {2J^3_L\over (p^0)^2})$, and vanishing
D4 charge $p^A=0$.

In section 3 we argue that an  appropriate BPS partition function
(i.e. index) $Z$ should not depend on the radius $R$, yielding an
equality of the form $Z_{4D}=Z_{5D}$ with a certain relation
between the arguments.  The microscopic description for many (but
not all) 5D spinning black holes is known \refs{\ascv, \bmpv}.
Hence this 5D-4D relation gives a microscopic description of 4D
black holes for many cases in which it had previously been
unknown. As a check these relations are found to correctly, and in
a rather intricate manner,  reproduce the entropy formula  at
leading order.

In section 4 we use this and a prior result \osv\ to give a
relation of the form\foot{ As discussed in \osv\ and \DabhBY\
\VerCK\ there are a number of subtleties in interpreting the 4D
version of this formula which of course also pertain here.}
\eqn\yuio{Z_{5D}(\mu)=|Z_{top}(g_{top}={8\pi^2\over \mu})|^2,}
between the BPS 5D black hole partition function and the
topological string partition function. Here $\mu$ is the potential
for $SU(2)_L$ spin and $g_{top}$ is the topological string
coupling constant. \yuio\ is quite different from the $linear$
relation of \refs{\gv,\kkv} between a certain 5D BPS partition
function and $Z_{top}$ and involving $g_{top} \sim \mu$. Combining
\yuio\ with the results of \refs{\gv,\kkv} potentially leads to a
non-trivial relation between $Z_{top}$ at different points in the
moduli space.

In section 5 the result is generalized to include general D4
charge $p^A$. From the 5D M-theory perspective this involves
turning on a four form $F^{(4)} \sim \omega_{NUT}\wedge
p^A\alpha_A$, where $ \omega_{NUT}$ is a harmonic Taub-NUT two
form and $\alpha_A$ is an integral  basis of harmonic Calabi-Yau
two forms. The 4D partition function for $any$ set of D-brane
charges may then be identified with that of a spinning 5D black
hole in this Taub-NUT-flux background. This identification is
again shown to intricately  yield the correct leading-order
entropy.

\newsec{M $\to$ IIA}

Consider $p^0$ D6 branes wrapping a Calabi-Yau space $X$ in a IIA
string compactification.  In the M-theory picture this is described
as the product of a Taub-NUT space with $X$: \eqn\taub{ ds^2_M =
\left(1+{p^0R\over r}\right)d\vec r^2  + R^2\left(1+{p^0R\over
r}\right)^{-1}(dx^{11}+ p_0 \cos \theta d\phi)^2 +ds_{X}^2-dt^2}
where $x^{11}\sim x^{11}+4\pi$. The Taub-NUT geometry has a
$U(1)_L\times SU(2)_R$ isometry, where the $U(1)_L$ generates
$x^{11}$ translations. The radius $R$ here is related to the
ten-dimensional IIA coupling via \eqn\cxv{R=g_{10}^{2/3}.}

At strong coupling, or large $R$, there is a large region with
$r\ll R$ in the core of the Taub-NUT geometry in which the 5D
metric reduces to \eqn\tab{ ds^2_{5} = { p^0 R \over r}(dr^2 + r^2
d\theta^2 + r^2 \sin^2 \theta d\phi^2 + r^2(dx^{11}/p_0 + \cos
\theta d\phi)^2) -dt^2 .} This is the flat metric on $R^4/Z_{p^0}$
tensored with the time direction. For $p^0=1$ we simply have 5D
Minkowski space.

Calabi-Yau compactifications of M theory to 5D  admit a second set
of supersymetric solutions with $U(1)_L\times SU(2)_R$ isometries.
These are the 5D spinning black holes \bmpv, characterized by
membrane charges $q_A$ and angular momentum $J_L$ associated to
the $U(1)_L$ isometry. Their characteristic size $r_{BH}$ grows as
the square root of the graviphoton charge $\sqrt Q$ which in turn
is proportional to the membrane charge $q_A$.

Let us now suppose that $\sqrt{Q}\ll R$ and $p^0=1$. Then we can
make an approximate BPS solution by inserting the spinning black
hole at the center of the $p^0=1$ Taub-NUT, symmetries aligned, well
inside the region where the $R^4$ is flat. Aligning the symmetries
requires the black hole to be exactly at the center of the Taub-NUT.
In fact an exact solution of this form exists for all $Q, ~R$
\gghpr\ and is reproduced in the appendix. Of course for large
$\sqrt{Q}\gg R$ it can no longer be described as a black hole in the
center of Taub-NUT, but this is irrelevant for our purposes since
the BPS quantities we consider should be independent of $R$.

 At distances large
compared to $R$, this solution is effectively a spherically
symmetric black hole in a four dimensional IIA compactification
carrying D6 charge $p^0=1$, and D2 charge $q_A$. In addition
 $J_L$, which is the eigenvalue
of  $U(1)_L$ rotations, becomes proportional to D0 charge $q_0$,
since $U(1)_L$ generates $x^{11}$ translations. To get the
proportionality factor, consider an orbit of the asymptotic
$U(1)_L$ in the $S^3$ near the tip ${\bf R}^4$. The angular
momentum in the 1-2 plane $J_1$ and that in the 3-4 plane $J_2$
are related to $J_L, J_R$ by $J_1=J_L+J_R, J_2=J_L-J_R$. An orbit
of the $U(1)_L$ is a helix going along a circle in the 1-2 plane
and a circle in the 3-4 plane at the same time. The wave function
of angular momentum $(J_L, J_R=0)$ picks up a factor $e^{2\pi
i(J_1+J_2)}=e^{4\pi iJ_L}$ as one goes around the $S^1$ orbit.
Therefore we conclude \eqn\ftm{q_0=2J_L.}

A similar construction works for integral $p^0>1$. We simply take
the exact 5D  solution and quotient it by the $Z_{p^0}$ subgroup
of the $U(1)_L$ isometry, which acts freely outside the horizon.
At infinity, this quotients the Kaluza-Klein circle and changes
its radius from $R$ to $R\over p^0$, while the topology of the 5D
horizon becomes $S^3/Z_{p^0}$. The corresponding 4D black hole
then has zero-brane charge\foot{Writing the D0 charge
schematically as a 4D spatial  integral $q_0 \sim \int d^4
\Sigma^b K^a T_{ab}$ of the $U(1)_L$ Killing field $K$ contracted
with the stress tensor, one factor of $p^0$ comes from the
division of the domain of the integrand, while the second comes
from demanding that $K$ be normalized so as to generate unit
translations of the Kaluza-Klein circle at infinity.}
\eqn\rtyh{q_0={2J_L \over (p^0)^2}.} Moreover, since the
$S^2\times S^1$ at infinity over which the 4D charges are given as
field strength integrals is divided by $p^0$, we have \eqn\tugh{
q_A={q_A^{5D}\over p^0}.}

\newsec{D6-D2-D0 Entropy}

The preceding classical construction suggests the quantum
conjecture that the supersymmetric partition function of a 4D
black hole with D-brane charges $(p^0,0,q_A,q_0)$ is directly
related to that of a $Z_{p^0}$ orbifold (which is trivial for
$p^0=1$) of a 5D black hole with membrane charges $q_A$ and spin
$q_0/2$. A precise conjecture relating certain 4D and  5D
supersymmetric indices will be made in the next section. In this
section we will check the conjecture at the level of the leading
semiclassical entropy.

The macroscopic entropy of a 5D spinning black hole is \kall\
\eqn\bmpven{ S_{5DBH} = 2\pi \sqrt{Q^3-J_L^2} ,} where $Q^3 =
D_{ABC}Y^AY^BY^C$ with $Y^A$'s satisfying $3D_{ABC}Y^BY^C=q_A^{5D}
$. A 4D black hole is obtained by inserting this 5D black hole in
the center of Taub-NUT. For the special case $p^0=1$, we identify
$J_L=q_0/2$, and \bmpven\ becomes \eqn\fxo{S_{4DBH}(p^0=1) =
2\pi\sqrt{ Q^3-{1\over 4}( q_0)^2}.} This agrees precisely with
the known 4D result for no D4 charges and $p^0=1$ \shmakova.

This is to be expected: in the reduction from 5D supergravity to
4D supergravity the radius of the fifth dimension is identified
with an appropriate combination of the 4D scalar moduli, and the
Taub-NUT radius is the asymptotic value of that scalar modulus at
infinity. The entropy of a 4D BPS black hole does not depend of
the asymptotic values of the scalar moduli at infinity.

Therefore, any microscopic accounting of a 5D black hole with
charges $q_A$ directly descends to a microscopic accounting of a
4D black hole with D6 charge $p^0=1$, D4 charge $p^A=0$, and
arbitrary D2-D0 charges $q_A, ~q_0$.

Now consider $p^0>1$. Dividing by $p^0$ divides the area and hence
the entropy by $p^0$. Therefore, in terms of the parameters $J_L$
and $Q_{5D}$ of the unquotiented 5D black hole the 4D entropy is
\eqn\fxo{S_{4DBH} = {2\pi\over p^0}\sqrt{ Q_{5D}^3-J_L^2}.} Using
\rtyh\ and \tugh\ then gives \eqn\ensft{ S_{4DBH} = 2\pi\sqrt{p^0
Q^3-{1\over 4}(p^0 q_0)^2} } in precise agreement with the 4D
entropy formula for general nonzero D0, D2 and D6 charges
\shmakova.

For $p^0>1$ a microscopic accounting of 5D entropy does not
descend so directly to an accounting of 4D entropy, because we
still have to understand the effect of the  $Z_{p^0}$ orbifold
action on the dual quantum microsystem describing the black hole.
The dual quantum microsystem is not known in general so we can't
describe the orbifold action. In order to proceed we assume a
microscopic picture of the kind discovered in \refs{\ascv,\bmpv},
in which the $U(1)_L$ corresponds to a conserved left-moving
current of a 2D CFT. $Z_{p^0}$ is then an orbifold action, and the
entropy is dominated by the ``long string" of the maximally
twisted sector. This effectively increases the 2D central charge
by a factor of $p^0$ so that $Q^3 \to p^0Q^3$. At the same time
the relation between worldvolume momentum and target space one is
rescaled as well $q_0\to p^0q_0$, and we recover \ensft. With this
assumption, any microscopic accounting of a 5D black hole with
charges $q_A$ directly descends to a microscopic accounting of a
4D black hole with D4 charge $p^A=0$, and arbitrary D6-D2-D0
charges $p^0,~q_A, ~q_0$. In section 5 we will relax the
restriction $p^A=0$.

\newsec{Spinning black hole and topological strings}
We conjecture the exact relation between the partition function of
4D extremal black holes and 5D spinning black holes, as follows
\eqn\ffdbh{ Z_{4D}(\phi^A, \phi^0) = Z_{5D}(\phi^A, 2\phi^0+2\pi i)
} where these partition functions are Witten indices of the form
\eqn\fdinx{ Z_{4D}( \phi^A, \phi^0) = {\rm Tr}'_{p^0=1, p^A=0}
(-1)^{2J^3} e^{-\phi^A q_A - \phi^0 q_0-\beta H} } and
\eqn\ffdbhinx{ Z_{5D}(\phi^A, \mu) = {\rm Tr}(-1)^{2J^3_L+2J^3_R}
e^{-\phi^A q_A - \mu J_L^3-\beta H} .} ${\rm Tr}'$ here denotes the
trace over all 4D states with the overall center-of-mass multiplet
factored out\foot{In 5D, this degree of freedom is part of the
background Taub-NUT geometry which is frozen.} and $J^3$ generates a
4D spatial rotation. The 4D trace is restricted to the sector with
$p^0=1$ and $p^A=0$. $Z_{5D}$ has IR divergences from black holes
which fragment and separate: we regulate these by putting them in
Taub-NUT space of radius $R$ which forces all black holes to sit at
the center (where they do not break supersymmetry),\foot{More
precisely, the quantum wave function of a hypermultiplet has one
supersymmetric ground state corresponding to the unique normalizable
self-dual harmonic two form $\omega_{NUT}$. An interesting
generalization, on which we hope to report, involves the
supersymmetric black ring.}and then taking $R\to \infty$. Using
$J^3(4D)=J^3_R(5D)$, $2q_0=J^3_L(5D)$ and the relation
$Z_{4D}=|Z_{top}|^2$ of \osv\ we have for $Z_{5D}$ \eqn\canon{
Z_{5D}(\phi^A, \mu=2\phi^0-2\pi i)=Z_{4D}( \phi^A, \phi^0 )=\left|
Z_{top}\left(g_{top}={8\pi^2 \over \mu }, t^A = {2 \pi \phi^A\over
\mu }\right) \right|^2 } Here $t^A$ are the K\"ahler moduli for the
topological string, $\phi^0$ is understood to be real.

This relation can be generalized  to  $p^0>1$ and/or $p^A\neq 0$
(see the next section) but additional assumptions are required.
\ffdbh\ seems to be the simplest of the relations between 4D and
5D black holes.

\newsec{The D6-D4-D2-D0 system}
In this section we generalize our construction to 4D extremal
black hole of generic charges $(p^0, p^A,q_A, q_0)$.

\subsec{$p^0=1$} In this subsection we take $p^0=1$ and then
generalize to $p^0>1$ in the next subsection. Consider turning on
a constant worldvolume $U(1)$ gauge field $F_{world}=p^A\alpha_A$
on a IIA $D6$ brane wrapping the Calabi-Yau $X$. The coupling of
$F_{world}$ to RR potential gives an object in 4D with charges
\eqn\hsz{(1, p^A,{3} p^A p^B D_{ABC}, -  p^A p^B p^C D_{ABC} ).}
Solving the attractor equations for such charges, we find simply
$CX^A = p^A , CX^0 = 1$ (see \aes\ for notation). The leading
order macroscopic entropy formula \aes\ then gives vanishing
entropy. This is consistent with the microscopic picture in which
there is a unique $F_{world}$.

Now let us try to understand the 11-dimensional description of
this configuration. The M-theory lift is again a Taub-NUT
geometry, with nonzero four form flux turned on: \eqn\iik{F^{(4)}
= \omega_{NUT} \wedge \sum p^A \alpha_A.}$\omega_{NUT}$ here is
the unique self-dual harmonic two form on Taub-NUT space
\GregoryTE, and $\alpha_A$ is a basis for harmonic Calabi-Yau
two-forms.  This flux sources D2 charge via the coupling $\int
C^{(3)}\wedge F^{(4)}\wedge F^{(4)},$ yielding $q_A= {3} p^B p^C
D_{ABC}$ as in \hsz. There is a nonzero Poynting vector
corresponding to the momentum along the M-theory circle. From the
4D point of view this is interpreted as D0 charge  $q_0=- p^A p^B
p^C D_{ABC}$ as in \hsz. So, by turning on $F^{(4)}$ as in \iik,
we produce a configuration with $p^0=1$, arbitrary D4 charges, but
predetermined D2-D0 charges and no entropy.

To get a configuration with arbitrary D2-D0 charges, we now insert
a 5D spinning black hole with charges $q^{5D}_{A}$ and angular
momentum $J^3_L$ in the middle of this Taub-NUT-flux
configuration. The exact solution can be found in \gghpr. This
yields a configuration with asymptotic 4D charges \eqn\ffg{(1,
p^A,{3 } p^A p^B D_{ABC} + q^{5D}_A, -  p^A p^B p^C D_{ABC} - {p^A
q_A^{5D}} + 2J^3_L )}  Notice the extra shift in $D0$ brane charge
coming from placing the charged 5D black hole in the nontrivial
magnetic four form field. This is a higher dimensional
generalization of Dirac's observation that a static electric
charge in a magnetic field carries angular momentum.

Now we wish to identify the partition function of the 4D black
hole with that of the spinning 5D black hole. 5D black holes
doesn't carry $p^A$ charge, so in order for this to be correct it
must be the case that, for the special values of charges given in
\ffg, the index $Z_{4D}$ is independent of $p^A$. This can be seen
as  a consequence of symplectic invariance, as follows.

The index $Z_{4D}$ is naturally a function of
$CX^\Sigma=p^\Sigma+i{\phi^\Sigma \over \pi}$ \osv. For a cubic
prepotential   $p^0=1$ and $p^A=0$, the electric potentials
$\phi^\Sigma$ are determined from the charges by \eqn\mhj{q_0=-Im
{C D_{ABC}X^AX^BX^C\over (X^0)^2}=Re{D_{ABC}\phi^A\phi^B\phi^C
\over \pi(\pi+i\phi^0)^2},~~~~}\eqn\fdz{q_A =3Im {C
D_{ABC}X^BX^C\over X^0}= -3 Im{D_{ABC}\phi^B\phi^C \over
\pi(\pi+i\phi^0)}. } Now consider the symplectic transformation
\eqn\hju{CX'^0= CX^0,~~~~~~~~~~~CX'^A= CX^A+{p^A}CX^0,} under
which $Z_{4D}$ is presumed invariant.\foot{In principle it might
transform as a modular form, but this would not affect the leading
order computation given here.} For the values of the moduli under
consideration this results in \eqn\fdy{X'^0=1+i{\phi^0 \over
\pi},~~ CX'^A=p^A+i({\phi^A \over \pi}+p^A{\phi^0\over \pi}).}
Comparing with \mhj\ we see that the new charges are related to
the old ones by \eqn\tphd{ q'_0=-Im { C D_{ABC}X'^AX'^BX'^C\over
(X'^0)^2}=q_0-p^Aq_A-D_{ABC}p^Ap^Bp^C,} and
\eqn\iod{q'_A=q_A+3D_{ABC}p^Bp^C.} Taking
$(q_A,q_0)=(q^{5D}_A,2J^3_L)$, this shift agrees exactly with that
encountered in \ffg. Therefore we can use a symplectic
transformation to shift from $p^A=0$ to arbitrary nonzero $p_A$
and $Z_{4D}$ remains unchanged. Physically this corresponds to the
fact that putting a 5D spinning black hole in a background
$F^{(4)}$ shifts some charges but does not change the number of
microstates. \subsec{$p^0>1$} A similar analysis holds for
$p^0>1$. The asymptotic charges \ffg\ for a spinning black hole
become \eqn\fsfg{(p^0, p^A,{3 \over p^0} p^A p^B D_{ABC} +
q^{5D}_A, -{1 \over (p^0)^2} p^A p^B p^C D_{ABC} - {p^A
q_A^{5D}\over p^0} + 2J^3_L ).} $p^A$ can then be shifted away as
before via the symplectic transformation \eqn\hjxu{CX'^0=
CX^0,~~~~~~~~~~~CX'^A= CX^A+{p^A\over p^0}CX^0.}

\centerline{\bf Acknowledgements} This work was supported in part
by DOE grant DE-FG02-91ER40654.

\appendix{A}{Supergravity solutions of spinning black hole in Taub-NUT space}
The Killing spinor equation of ${\cal N}=2$ 5D supergravity is
\eqn\killing{ \left[ d+ {1\over 4}\omega_{ab} \Gamma^{ab} + {i\over
4\sqrt{3}} e^a ({\Gamma^{bc}}_a F_{bc} - 4 \Gamma^b F_{ab}) \right]
\epsilon =0 } where $e^a$ are the frame 1-forms and $\omega_{ab}$ is
the spin connection. The metric for the supersymmetric spinning
black hole in Taub-NUT space is \gghpr\ \eqn\metric{ ds^2 =
-(1+{\tilde Q\over Rr})^{-2} \left(dt +{\tilde Ja\over p^0R^2}
\right)^2 + (1+{\tilde Q\over Rr}) ds_{TN}^2 } where \eqn\adef{
a=\left(1+{p^0R\over r}\right)(dx^{11}+ p^0 \cos \theta d\phi) -
dx^{11} } and $x^{11}\sim x^{11}+4\pi$. $R$ is the asymptotic radius
of the Taub-NUT space and the graviphoton field \eqn\gravi{ F =
{\sqrt{3}\over 2} d \left[(1+{\tilde Q\over Rr})^{-1} \left(dt
+{\tilde Ja\over p^0R^2} \right) \right]} Similarly to the
calculation of \gmt\ , the Killing spinor equations are solved by
\eqn\killcond{ i\Gamma^0\epsilon = \epsilon } and the self-duality
of $da$ and of the spin connection of Taub-NUT space.

With a redefinition of variable $r=\rho^2/R$, in the limit $R\to
\infty$, the solution \metric\ becomes \eqn\rinfl{ ds^2 = -(1+{
\tilde Q\over \rho^2})^{-2} \left[dt + {\tilde J\over
\rho^2}(dx^{11}+p^0\cos\theta d\phi)\right]^2 +4p^0 (1+{\tilde
Q\over \rho^2}) (d\rho^2 + \rho^2 d\tilde \Omega_3^2) } where
\eqn\sthem{ d\tilde \Omega_3^2 = {1\over 4} \left[ d\theta^2
+\sin^2\theta d\phi^2 + {1\over (p^0)^2}(dx^{11}+\cos\theta d\phi)^2
\right] } is the metric on the unit $S^3/{\bf Z}_{p^0}$. \rinfl\ is
nothing but a spinning black hole at the center of the orbifold
space ${\bf C}^2/{\bf Z}_{p^0}$. Note that the area of the black
hole horizon is independent of $R$, and is given by \eqn\areah{ A =
{16 \pi^2} \sqrt{p^0 \tilde Q^3 - (p^0 \tilde J)^2} } $\tilde Q$ and
$\tilde J$ are related to the standard normalized 5D charges $Q,J$
\bmpv\ by a rescaling, \eqn\rescla{ Q = 2\pi^{2/3} \tilde Q, ~~~J =
2\sqrt{2}\pi \tilde J. }

\listrefs

\end
We assume that under symplectic transformations $M$ that $Z_{4D}$
is a
scalar\eqn\jikd{Z_{4D}CX^\Lambda)=S_{BH}(M^{\Lambda}_{\Sigma}CX^\Sigma).}
This implies that the charge

SO we want to show that $Z_{4D}(1+i{\phi^0\over \pi}, i{\phi^A
\over \pi})$  and $Z_{4D}(1+i{\phi^0 '\over \pi},p^A+ i{\phi^A '
\over \pi})$, as long as \eqn\rtl{q_0=-{\p \over \p \phi^0}\ln
Z_{4D}(1+i{\phi^0

\over \pi}, i{\phi^A \over \pi})...}

\eqn\mbn{S_{4D}( p^0,p^A,q_A = {3\over p^0} p^A p^B D_{ABC} +
\tilde q_A, q_0 =  - {1\over(p^0)^2} p^A p^B p^C D_{ABC} - {p^A
\tilde q_A\over p^0} + \tilde q_0) = S_{4D}(p^0,0,\tilde q_A,
\tilde q_0).}

This is indeed true. It can be seen directly from the entropy
formula, but it is instructive to look at the entropy as the
extreme value of the near horizon central charge $|Z|^2 = |p^A F_A
- q_A X^A|^2$. For the usual tree level prepotential ( and even
including one-loop corrections) it is immediate to see that

\eqn\jka{Z(p^0,p^A,q_A = {3\over p^0} p^A p^B D_{ABC} + \tilde
q_A, q_0 = - {1 \over (p^0)^2} p^A p^B p^C D_{ABC} - {p^A \tilde
q_A \over p^0} + \tilde q_0, X^A,X^0) = Z(p^0,0,\tilde q_A, \tilde
q_0, X^A - {p^A \over p^0}X^0,X^0).} Hence the extrema coincide
for the original charges and the shifted ones.

\eqn\hsz{(p^0, p^A,{3\over p^0} p^A p^B D_{ABC}, - {1\over(p^0)^2}
p^A p^B p^C D_{ABC} ).} \subsec{ Another way to say it} The black
hole entropy is a symplectic invariant function of the complex
coordinates \eqn\jjh{S_{BH}=(C), ~~~S_{BH}(} where $M$ is a
symplectic transformation.  This allows us to relate the entropy
of a BH hole with no D4 charge to one an arbitrary one with
nonzero D4 charge $p^A$. In particular considering the
transformation and using \jjh\ one finds
\eqn\ssa{S_{BH}(p^0+i{\phi^0 \over \pi},~~i{\phi^A \over
\pi})=S_{BH}(p^0+i{\phi^0 \over \pi},~~p^A+i{\phi^A \over
\pi}+i{\phi^0\over p^0}p^A).}  Using the cubic prepotential and
the formulae for $q_\Lambda$ as a function of the periods then
reproduces \mbn.